\documentclass[]{rmf-d}
\usepackage{nopageno,
multicol,times,amsmath,amssymb} 
\usepackage[english]{babel}   
\usepackage[]{caption2}
\usepackage{graphicx}
\def\rmfcornisa{ENSE\~NANZA \hfill\rmf\ {\bf 48} (6) ???--??? 
\hfill Febrero 2003}

\def\rmfcintilla{{\it Rev.\ Mex.\ F\'{\i}s.\/} {\bf 48} (6) (2003) ???--???}
\clearpage \rmfcaptionstyle 
\newcommand{\vev}[1]{\langle#1\rangle}
\begin{document}
\title{Heavy Meson Physics: What have we learned in Twenty Years?
\vspace{-6pt}}

\author{Benjam\'\i{}n Grinstein}
\address{University of California, San Diego; La Jolla, CA 92093, USA}

\maketitle \recibido{ de  de }{ de  de \vspace{-12pt}} 

\begin{abstract}
I give a personal account of the development of the field of heavy
quarks. After reviewing the experimental discovery of charm and
bottom quarks, I describe how the field's focus shifted towards
determination of CKM elements and how this has matured into a
precision science.
\end{abstract}
\begin{resumen}
Doy una perspectiva personal del desarrollo de la disciplina de quarks
pesados.  Despu\'es de repasar la historia del descubrimiento de los
quarks $c$ y $b$, doy una descripci\'on de como la disciplina se ha
re-enfocado hacia la determinaci\'on de los elementos de la matriz CKM y
de como ha madurado hasta convertirse en una ciencia de precisi\'on. 
\end{resumen}
\keys{Heavy Quarks, Effective Theory, Semileptonic Decays,
  Kobayashi-Maskawa Matrix \vspace{-4pt}} 
\pacs{14.40.Nd, 12.39.Hg, 13.20.He, 14.20.Mr, 13.25.Ft, 12.15.Hh,  \vspace{-4pt}}
\begin{multicols}{2}
\section{Introduction}
Let me start with a warning: this is {\it not} a review of heavy quark
physics. I have been honored by the Divisi\'on de Part\'\i culas y Campos of
the Sociedad Mexicana de F\'\i sica with the Medalla 2003, for which I am
grateful and humbled. This talk was given on occasion of the medal
being conferred. I tried to describe, from my personal view point, how
the field of $B$ physics evolved since its inception to the
present, biased by my own experience.

\section{Ancient history}

\subsection{While I was in High School \ldots}
In November of 1974 two experimental collaborations announced the
discovery of a new very narrow resonance with mass 3.1~GeV. They had
unearthed evidence for the charm quark, and  for the validity
of an asymptotically free theory, like QCD, for strong
interactions. These events had extraordinary consequences, affecting
the way we think today about particle physics. They are often
referred to as the ``November Revolution''.
 
A MIT--Brookhaven collaboration, led by S.~Ting, found evidence for
the new resonance by measuring the $e^+e^-$ mass spectrum in $p+{\rm
Be} \to e^++e^-+X$ with a precise pair spectrometer at Brookhaven
Natl.\ Lab.'s 30~GeV AGS\cite{jpsia}.

The Mark~I collaboration, from SLAC and LBL, led by B. Richter was
conducting experiments at the newly constructed $e^+e^-$ ring, SPEAR,
at SLAC. Their detector consisted of a spark chamber embedded in a
solenoidal magnetic field, and surrounded by time-of-flight counters,
shower counters and proportional counters embedded in slabs of
iron for muon identification. They\cite{jpsib} ``observed a very sharp
peak in the cross sections for $e^+e^-\to$~hadrons, $e^+e^-$, and
$\mu^+\mu^-$ at a center of mass energy of $3.105\pm0.003$~GeV'' and
found an upper bound on the width of 1.3~MeV.

The MIT-BNL collaboration called the new resonance ``J'', Mark~I
called it ``$\psi$'', so it's now known as the $J/\psi$. The Mark~I
collaboration soon found a second narrow resonance\cite{psip}, the
$\psi'$, at a mass of 3.695$\pm$0.004~GeV. No other narrow resonances
were found in the total $e^+e^-$ cross sections at SPEAR, but broader
structures did appear at energies above the
$\psi'$\cite{siegrist}. Other narrow structures that could not be
directly produced in $e^+e^-$ collisions were found through cascade
decays of the $\psi'$. The DASP collaboration working at DESY's
$e^+e^-$ storage ring DORIS found\cite{dasp} the first $\chi$ state in
$\psi'\to\chi+\gamma\to\psi+\gamma+\gamma$. The Crystal Ball
collaboration\cite{crysball} detector provided the high spatial and
energy resolution needed to finally unravel the spectroscopic levels
of charmonium.  

The interpretation of these resonances soon became clear: they are
atom-like bound states of a charm quark-antiquark pair. The
interaction between the rather heavy quarks is coulomb like, since at
short distances QCD becomes weak and a single gluon exchange gives an
attractive coulomb potential between the quarks. The potential is not
really coulomb: for one thing, it must be confining so it must grow
without bound at long distances. But the physics of the spectrum of
bound states is dominated by the short distance interaction and is not
dissimilar from the physics of the hydrogen atom. The $\eta_c$ and
$\psi$ families are in a quark-spin singlet and triplet state,
respectively, with orbital angular momentum~0, giving $J^{PC}=0^{-+}$
and $1^{--}$, respectively. The $\chi_{c0}$, $\chi_{c1}$ and
$\chi_{c2}$ are in a spin triplet, with $J^{PC}=0^{++}$, $1^{++}$ and
$2^{++}$, respectively.

Charmonium states have zero charm number, $C$. States with $|C|=1$, with so
called ``naked charm'', were first convincingly observed by Mark~I at
SPEAR\cite{dmsnmrki}. They observed narrow peaks in the invariant mass
spectra for neutral combinations of charged particles in $K\pi$ and
$K3\pi$. They inferred the existence of an object of mass
$1865\pm15$~MeV and put an upper limit on its width of 40~MeV. 
 The new state, with $C=1(-1)$, was the $D(\bar D)$
pseudoscalar meson. They found ``it significant that the threshold
energy for pair-producing this state lies in the small interval
between the very narrow $\psi'$ and the broader \dots''
$\psi''$. That is,  the $\psi''$ is much broader
because it decays strongly into a $D$--$\bar D$ pair.

\subsection{\ldots and then in college}
The discovery of ``naked bottom'' (or ``naked beauty'', outside the
Americas) paralleled in many ways that of charm. Although a new
sequential heavy lepton, the $\tau$, had been discovered, and
therefore the existence of beauty and top expected, the masses of
these quarks were unknown.

L.~Lederman led a collaboration at Fermilab that used a two arm
spectrometer to search for muon pairs in 400~GeV proton-nucleus
collisions. They had some experience. Years earlier the group
conducted a similar experiment at BNL's AGS. Because their apparatus
had smaller resolution than that of the MIT-BNL group, they did not
report any evidence for a resonance. They had seen a cross section
that, except for a small plateau in the 3~GeV region, fell with
invariant mass as expected. After missing the $J/\psi$, they were ready
for the discovery of bottomonium. They observed\cite{upsifermi} a
similar effect in the new experiment, and correctly interpreted it as
a dimuon resonance at about 9.5~GeV. A refined analysis of the
experiment revealed actually two peaks, at 9.44 and 10.17~GeV. The
states were named ``$\Upsilon$'' and ``$\Upsilon'$''.

An upgrade of the energy of DORIS made it possible for the PLUTO and
DASP~II collaborations to observe the $\Upsilon$ in $e^+e^-$
annihilation\cite{upsipluto,upsidasp}. A
further energy upgrade made the $\Upsilon'$ accessible
too\cite{upsipdasp,upsipdesyh}.

After the Cornell Electron Storage Ring (CESR) was commissioned, the
CUSB and CLEO collaborations successfully observed the $\Upsilon$,
$\Upsilon'$ and $\Upsilon''$. All three resonances, with masses 9.460,
10.023 and 10.355~GeV are narrow.  Shortly afterwards the two
collaborations established the existence of a broader resonance, the
$\Upsilon'''$, at a mass of $\sim10.55$~GeV and a width of
about~$12.6$~MeV.  This is significant because, following the charm
experience, it suggests looking for naked beauty in the decay of
$\Upsilon'''$.  $B$-mesons were first found and reported by the CLEO
collaboration in a paper which for once is straight and to the point
in its title (``Observation of Exclusive Decay Modes of $b$-Flavored
Mesons'') and in its abstract (see Ref.~\cite{bmesoncleo}). To be sure,
$B$-mesons had been inferred from the observation of high momentum
leptons in $\Upsilon'''$ decays, but it was the reconstruction of a
few exclusive decays that demonstrated their existence conclusively.
Today $D$ and $B$ mesons are  universally accepted
established resonances. They are the closest we can get to having
naked charm and beauty. Their masses have been measured to high
accuracy\cite{Eidelman:2004wy}:
\begin{align}
m_{D^\pm}&=1869.4\pm0.5~{\rm MeV}\\
m_{D^0}&=1864.6\pm0.5~{\rm MeV}\\
m_{B^\pm}&=5279.0\pm0.5~{\rm MeV}\\
m_{B^0}&=5279.4\pm0.5~{\rm MeV}
\end{align}

\section{Lifetimes, CKM texture, Semileptonic and ISGW}
One of the first surprises encountered in the early 80's was the long
lifetime of naked $B$. The 1984 PDG Review of Particle
Properties\cite{pdg84} gives:
\begin{align}
\tau_{D^\pm}&=0.92\pm0.15~{\rm ps}\\
\tau_{D^0}&=0.44\pm0.07~{\rm ps}\\
\tau_{B}&=1.4\pm0.4~{\rm ps}
\end{align}
The $B^\pm$ and $B^0,\bar{B}^0$ lifetimes are not separated. The
$B$ lifetime is {\it longer} than the $D$ lifetime. Since naively
$\tau_D:\tau_B = (m_b/m_c)^5 (x|V_{cb}|^2+|V_{ub}|^2)/|V_{cs}|^2$, where $x\approx0.5$ is a
phase space suppression factor, one has to conclude that both $V_{cb}$
and $V_{ub}$ are small:
\begin{equation}
\sqrt{x|V_{cb}|^2+|V_{ub}|^2}\approx 
\left(\frac{m_c}{m_b}\right)^{\frac52}|V_{cs}|\sqrt{\frac{\tau_D}{\tau_B}}
\approx 0.03 - 0.05.
\end{equation}
Moreover, already the 1984 PDG Review listed the decay branching
fraction into charmed final states as $80\pm28$\%, implying that
$|V_{ub}|\ll |V_{cb}|$ so that the above estimate gives $|V_{cb}|\approx
0.05-0.07$. Unitarity of the CKM matrix requires $|V_{ub}|^2+
|V_{cb}|^2+|V_{tb}|^2=1$, so we learned, in addition, that $|V_{tb}|
\approx1$. There was a quantum leap in understanding of the texture of
the CKM matrix: it became evident that it has 1's along the
diagonal, numbers of order 0.1 off the diagonal, and very likely much
smaller than 0.1 two steps removed from the diagonal. The rule of
thumb was (and still is) $V_{ij}\sim (0.1)^{|i-j|}$. In what used to be
the standard parametrization of the CKM matrix in terms of angles
$\theta_i$ and $\delta$, this meant that all angles are of the order of
0.1. The origin of this texture remains a mystery, and a challenge to
model builders.

\begin{center}
\includegraphics[height=1.65in]{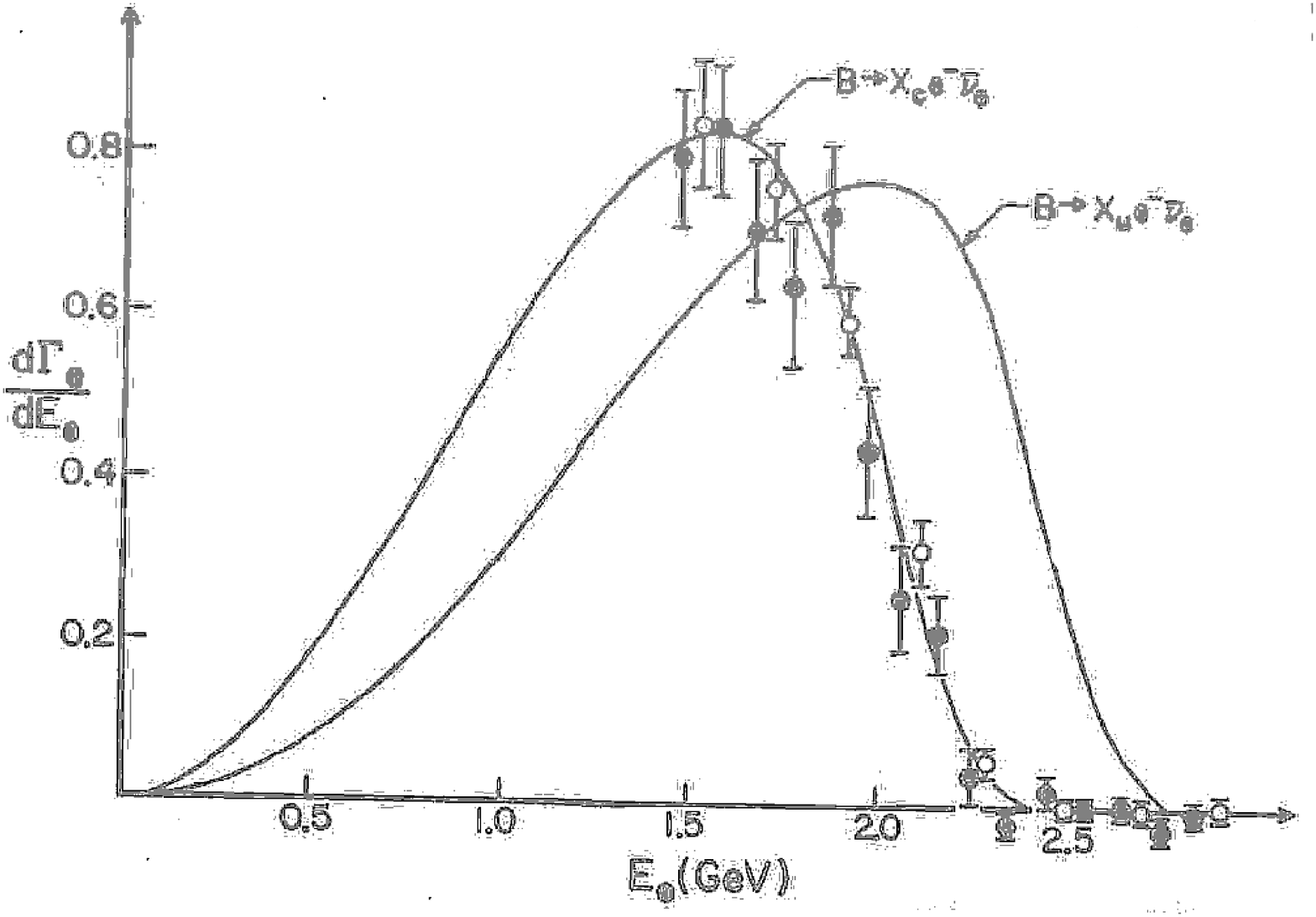}\\
Fig.~1
\end{center}

How can we progress from an estimate of the CKM angles to a precision
measurement? This is an example of a story that repeats itself over and
over in the study of heavy quarks: experiment and theory have to work
together to find a route to the answer of this
question. Experimentally it would be easiest to measure with great
precision the lifetime of the $B$ meson, but theorists did not have
(nor do they have today) a complete theory of lifetimes. Theory prefers
inclusive widths of semileptonic decays, but experimental backgrounds
make this a tough, if not impossible measurement, particularly when
restricting to non-charm final states as is necessary for the
determination of $|V_{ub}|$. The difficulty with the charm background 
is easily grasped from Fig.~1, which  shows data from
CLEO\cite{Stone:1983yy} and ARGUS\cite{Schroder:1985eq} superimposed
on model calculations (see below) of the semileptonic decay spectrum
into c- and u-quarks, assuming $V_{ub}=V_{cb}$. 

\begin{center}
\includegraphics[height=1.5in]{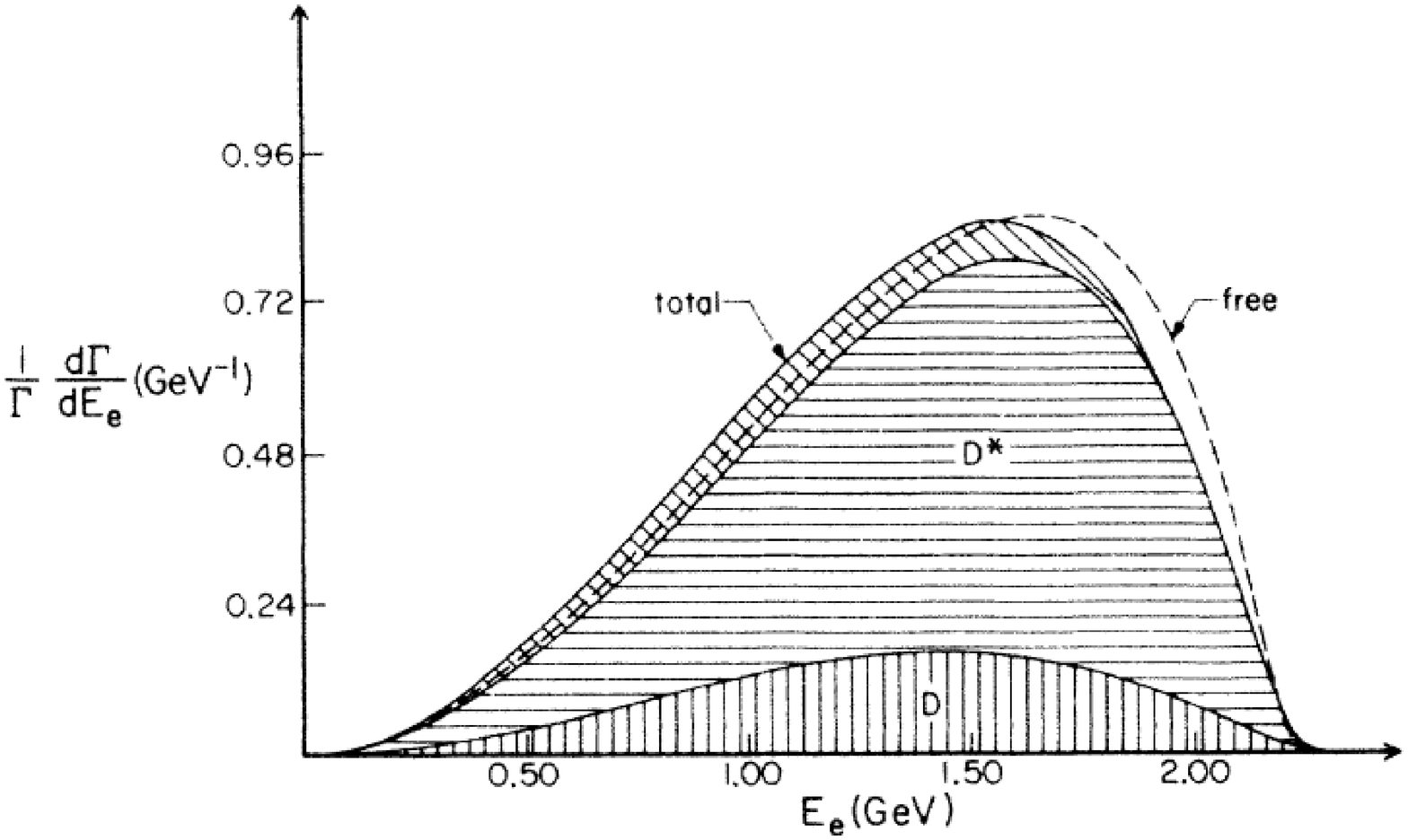}\\
Fig.~2
\end{center}

Not knowing how to calculate the decay rate from first principles,
theorists resorted to reasonable guesses, or ``hadron models.'' The
GSW model\cite{Grinstein:1986ad}, and later incarnations as the
ISGW\cite{Isgur:1988gb} and ISGW-II\cite{Scora:1995ty} models, was
simply an application of the quark-potential model of hadrons to the
computation of matrix elements of $V-A$ charged currents between an
initial $B$ meson state and a final state consisting of a single meson
with either a charm quark (for $V_{cb}$) or a u-quark (for
$V_{ub}$). It was reassuring that adding over a few final state
charmed mesons, the semi-inclusive decay rate into charm gives
approximately the same answer as the free quark decay rate. Fig.~2
shows how the individual charmed resonances add up. 

\begin{center}
\includegraphics[height=2in]{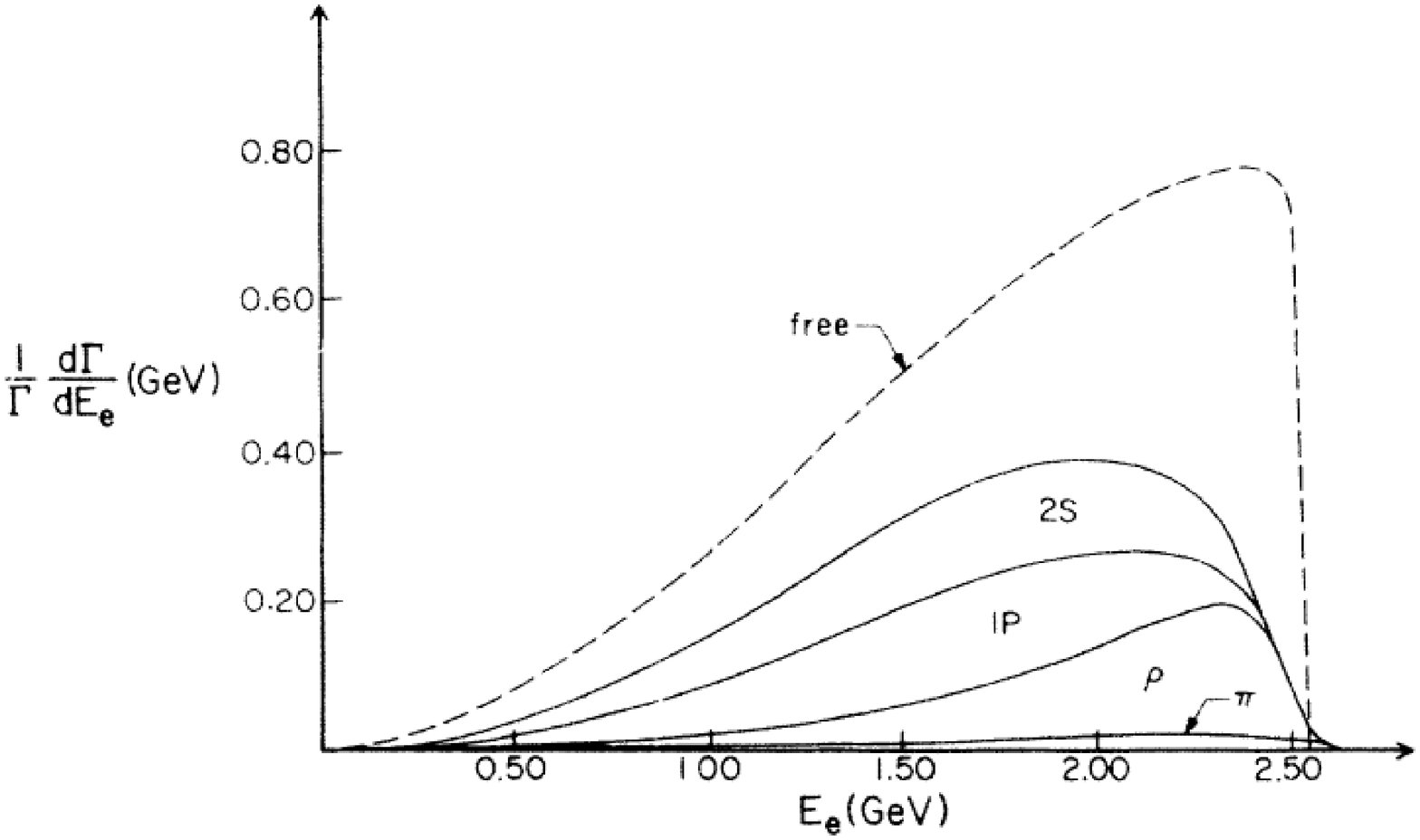}\\
Fig.~3
\end{center}

In order to determine $V_{ub}$ the charm background has to be
controlled. A favorite experimental method was to measure the
inclusive semileptonic decay spectrum as a function of the readily
measurable electron (or muon) energy, and  to focus on energies large
enough that decay into charm is forbidden. The calculation of GSW,
Fig.~3, showed that the rate in the restricted region is dominated by
a few final state resonances which are only a fraction of the total
semi-inclusive rate as calculated directly at the parton level ($b\to
ue\bar \nu $). This suggested that the parton level calculation
is highly unreliable in the end-point region. Moreover, the result of
summing over a few resonances had a different shape and end-point than
the partonic result. And, as if not enough, the model computation was
fairly sensitive to the choice of model parameters. Ouch!

\begin{center}
\includegraphics[height=1.2in]{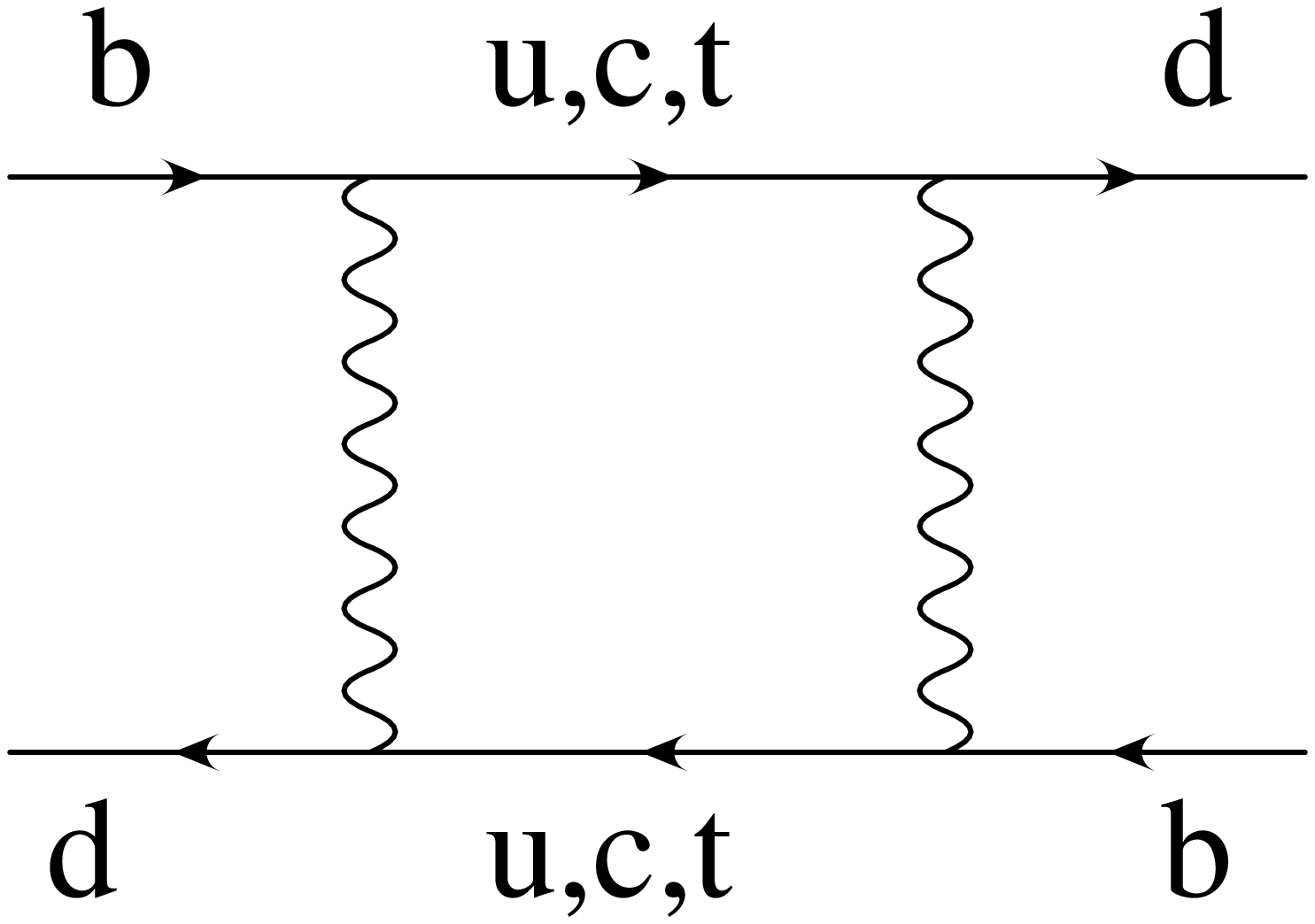}\\
Fig.~4
\end{center}

\section{Mixing, Heavy Top, Rare Decays}
The standard model predicts $B^0-\bar{B}^0$ oscillations, much in
the same way as for $K^0-\bar{K}^0$. The underlying weak process
is well known. Since mixing requires a change of $b$-number by two
units, there must be two $W^\pm$ exchanged, and the process is
doubly-weak. The amplitude is given by the ``box'' Feynman diagram of
Fig.~4.  Keeping track of the CKM factors from both fermion lines
in the box diagram, one has
\begin{equation}
\label{eq:box1}
{\rm Amp}\propto
G_F^2\!\!\!\sum_{i,j=u,c,t}\!\!\!(V_{ib}^{\phantom{*}}V^*_{id})(V_{jb}^{\phantom{*}}V^*_{jd})
F(m_i^2/M_W^2,m_j^2/M_W^2),
\end{equation}
where $F$ is a function that arises from computing the one loop graph.
If the three intermediate quarks were degenerate the amplitude would
vanish identically since the CKM matrix is unitary:
$\sum_{i=u,c,t}V_{ib}^{\phantom{*}}V^*_{id}=0$. Expanding $F$ in the quark masses we have,
\begin{equation}
\label{eq:box2}
{\rm Amp}\propto G_F^2\left(\sum_{i=u,c,t}\!\!V_{ib}^{\phantom{*}}V^*_{id}\frac{m_i^2}{M_W^2}\right)^2,
\end{equation}
Since each of the factors $V_{ib}^{\phantom{*}}V^*_{id}$ involves a jump of two
generations, the rule of thumb of the previous sections says that all
these are similar (and of order $(0.1)^2$), 
\begin{equation}
|V_{ub}^{\phantom{*}}V^*_{ud}|\sim|V_{cb}^{\phantom{*}}V^*_{cd}|\sim|V_{tb}^{\phantom{*}}V^*_{td}|.
\end{equation}
It follows immediately that the top quark gives the dominant
contribution  to the mixing amplitude in Eq.~(\ref{eq:box2}).  

In the absence of mixing, a $B^0-\bar{B}^0$  pair produced in
an $e^+e^-$ collision will produce opposite charged leptons when both
$B$'s decay semileptonicaly. Mixing implies that some fraction of the
time  the two semileptonic decays produce the same charge leptons.The  
ARGUS collaboration discovered this phenomenon and
reported\cite{Albrecht:1987dr}
\begin{equation}
r\equiv \frac{N(B^0B^0)+N(\bar{B}^0\bar{B}^0)}{N(B^0 {\bar B}^0
)}=
0.21\pm0.08.
\end{equation}
The rate of mixing is given by the amplitude above times $f_B^2B_B$,
which characterizes the matrix element, times some fixed numbers
(including short distance QCD corrections). To explain this rather
large mixing within the standard model of electroweak interactions we
could assume $f_B^2B_B$ was larger than estimated. But even taking
rather extreme values for  $f_B^2B_B$ (corresponding to more than four times the
modern accepted value!), and taking $|V_{tb}^{\phantom{*}}V^*_{td}|$ as large as
possible,  we were forced to require a large top quark mass, in excess
of 50~GeV.

\begin{center}
\includegraphics[height=1.2in]{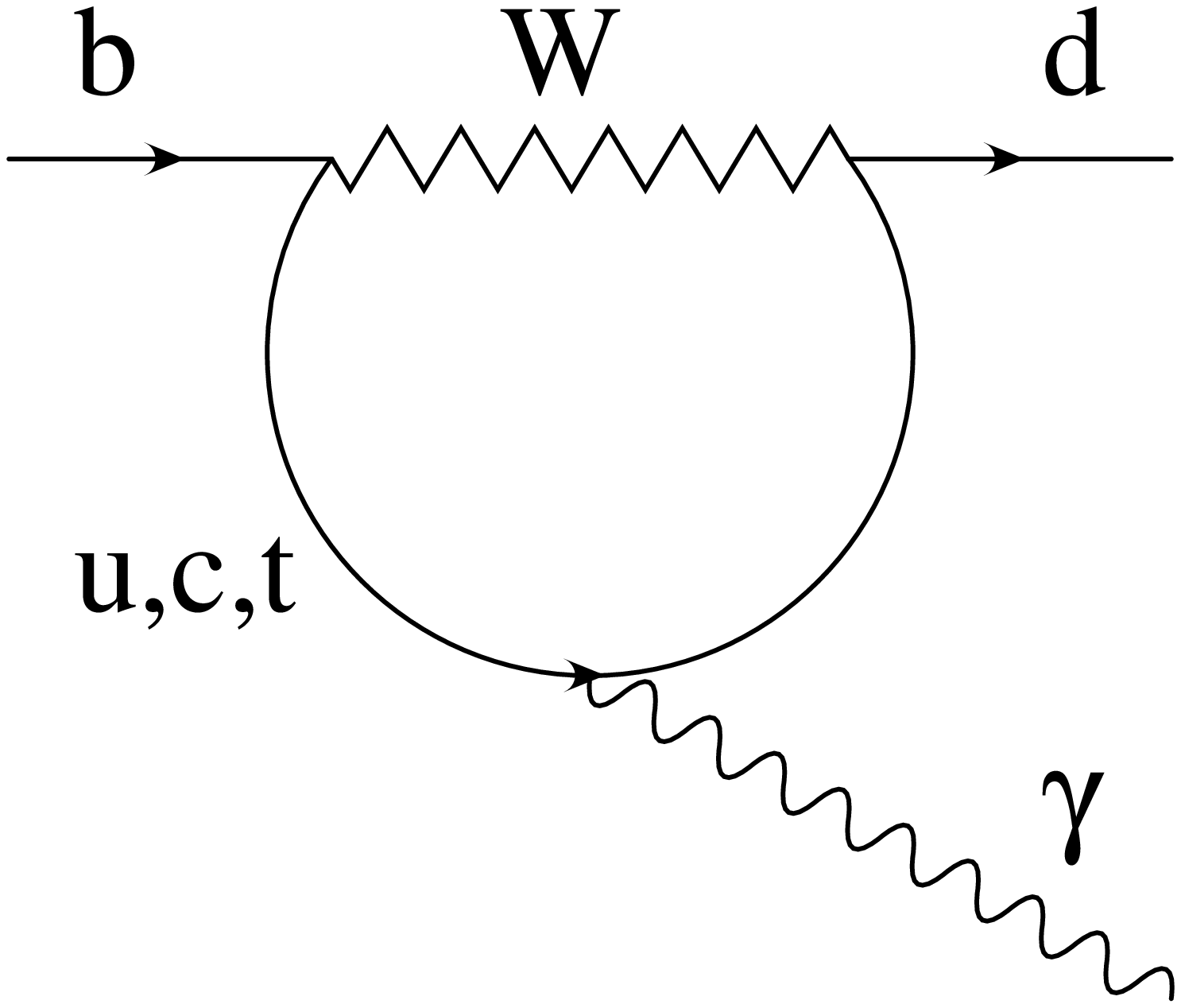}\\
Fig.~5
\end{center}

In 1987 the direct bound on the mass of the top quark was 15~GeV. It
was expected that the $t$-quark would be much lighter than the $W$, a
reasonable guess given the masses of all other quarks. The evidence
from $B^0-\bar{ B}^0$ mixing was that the top quark was much
heavier, possibly heavier than the $W$. This had immediate, surprising
implications.

For one, the GIM mechanism is a bit of a fluke. Take for example the
prediction of the mass of the charm quark. GIM cancellations bring in
a suppression factor of $m_c^2/M_W^2$, as can be seen by adapting the
result in Eq.~(\ref{eq:box2}) to the case of $K^0-\bar{ K}^0$
mixing. The top quark contribution is not suppressed by $m_t^2/M_W^2$
simply because this factor exceeds unity. In fact, for the top quark
it makes no sense to approximate the function $F$ in (\ref{eq:box1})
by the expansion in (\ref{eq:box2}). The top contribution to
$K^0-\bar{K}^0$ mixing involves $V_{td}^{\phantom{*}}V_{ts}^*$. This is a three
generation jump, as compared to the single jump for the charm
contribution, so we get a suppression factor of $(0.1)^2$. But, by
comparison, the charm contribution is suppressed by
$m_c^2/M_W^2\sim0.0004$.

\begin{center}
\includegraphics[height=1.7in]{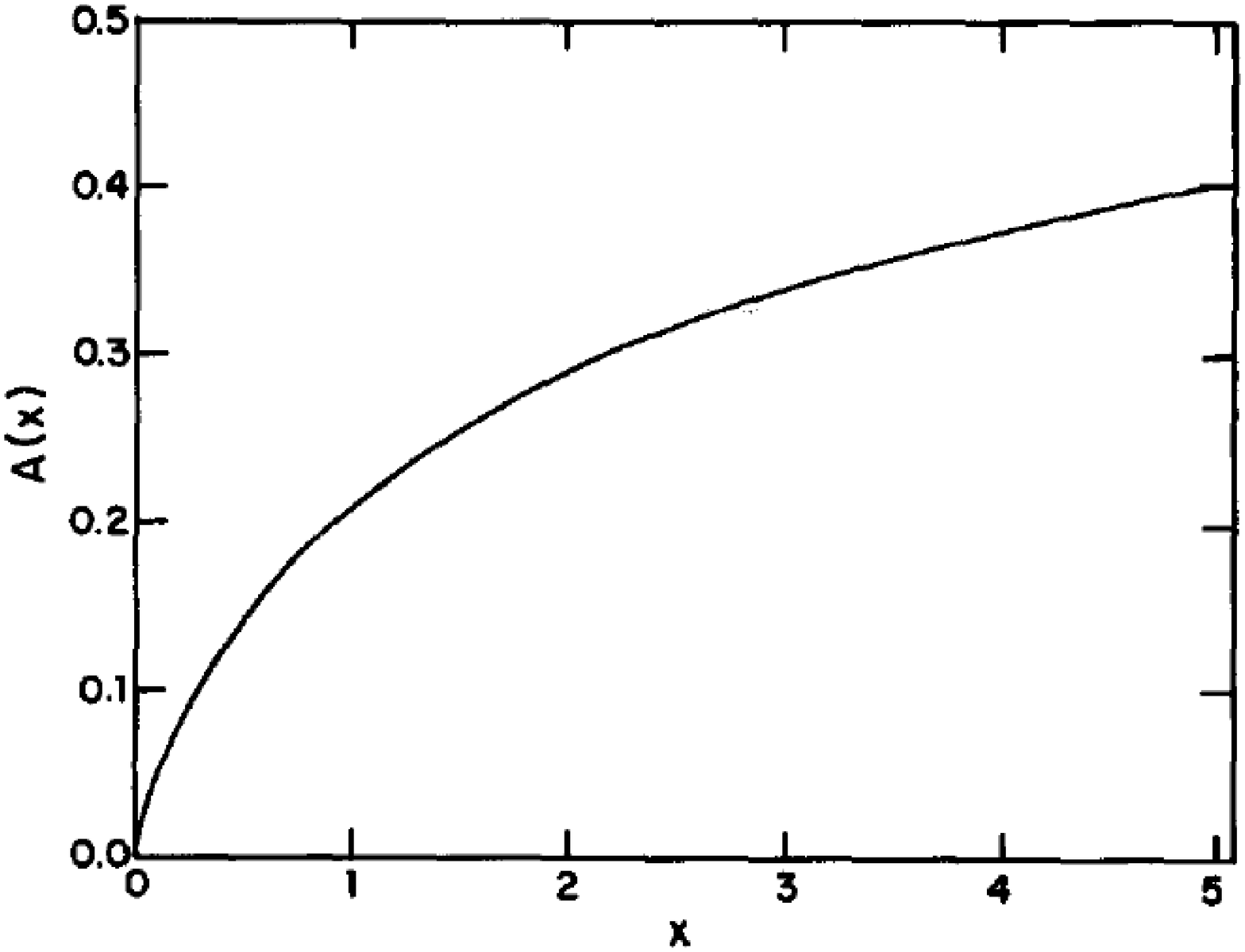}\\
Fig.~6
\end{center}

A second implication is that processes mediated by virtual top-quarks
may have larger rates than we had thought by 1987. The first such process
you would think of is the radiative decay, $b\to s\gamma$. The lowest
order Feynman diagram is the one loop graph shown in Fig.~5. This gives
an effective Hamiltonian for the radiative decay
\begin{equation}
\label{eq:heffbtosgamma}
\mathcal{H}_{\rm eff}=\frac{G_F}{\sqrt2}V_{tb}^{\phantom{*}}V^*_{ts}\frac{e
  m_b}{16\pi^2}A(\bar s_L\sigma^{\mu\nu}F_{\mu\nu}b_R).
\end{equation}
Here we have neglected the contribution of $u$ and $c$ quarks, and $A$
is a function of the top quark mass plotted in Fig.~6 as a function of
$x=m_t^2/M_W^2$. 

Short distance QCD corrections enhance the amplitude by about 70\%, so
the rate is enhanced by a full factor of three; see Fig.~7. The reason
the QCD corrections are so large is that the function $A$ in
(\ref{eq:heffbtosgamma}) is accidentally small. The QCD corrections
are suppressed by the strong coupling constant but enhanced by large
logarithms. Resumming large logarithms of the ratio of $m_b$ to $M_W$
or $m_t$ (assuming $m_t$ is the same scale as $M_W$) has the effect of
replacing $A$ in (\ref{eq:heffbtosgamma})\cite{Grinstein:1987vj}:
\begin{multline}
A\to\left(\frac{\alpha_s(m_b)}{\alpha_s(M_W)}\right)^{-\frac{16}{23}}
\left\{ A+ \frac{3a}{10}\left[
\left(\frac{\alpha_s(m_b)}{\alpha_s(M_W)}\right)^{\frac{10}{23}}-1\right]
\right.\\
\left. + \frac{3a}{28}\left[
\left(\frac{\alpha_s(m_b)}{\alpha_s(M_W)}\right)^{\frac{28}{23}}-1\right]
\right\}.
\end{multline}
Here $a=232/81$ ($-140/81$ for the case of $c\to u \gamma$). 

\begin{center}
\includegraphics[height=1.7in]{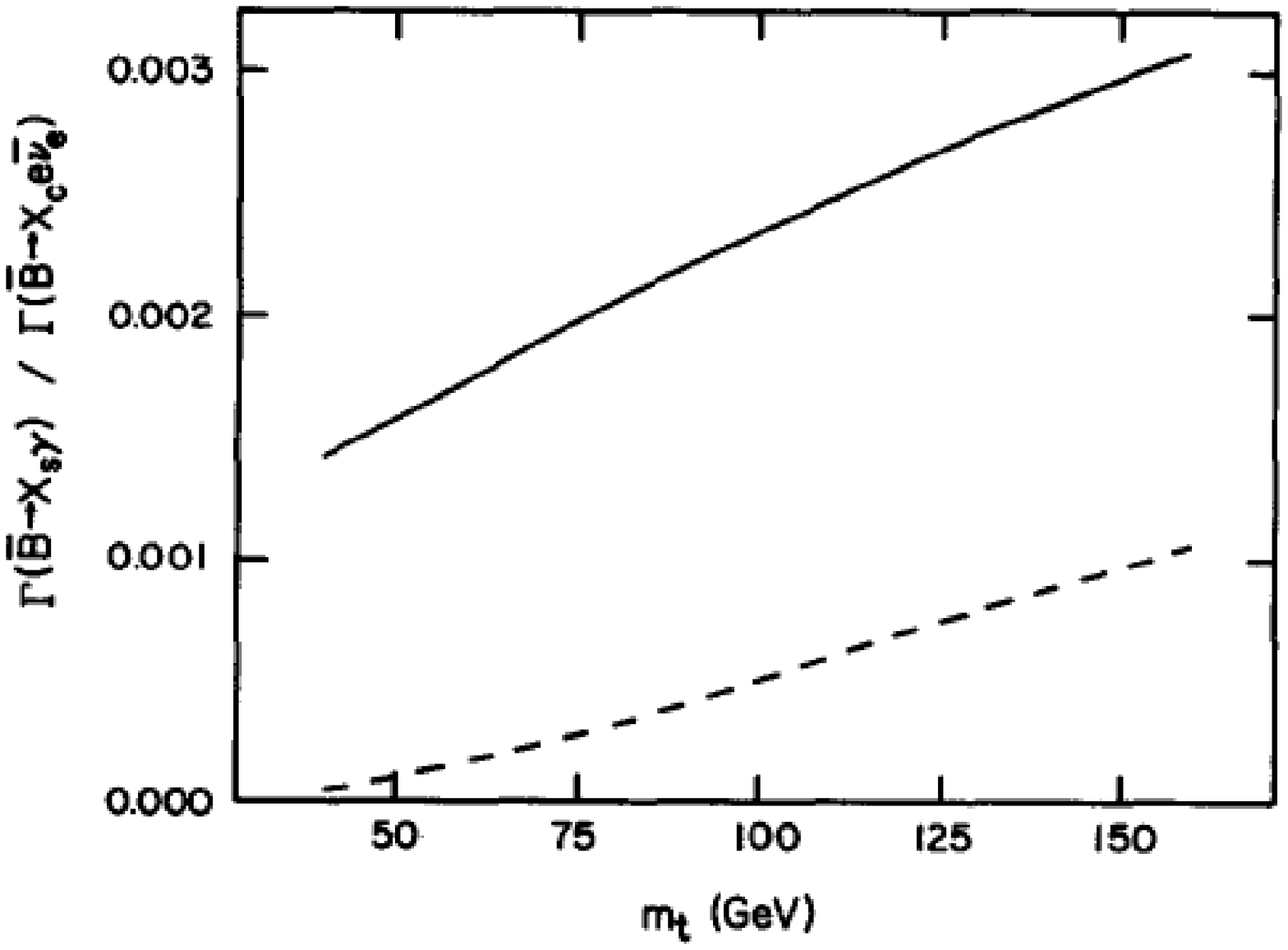}\\
Fig.~7
\end{center}

This is another interesting GIM fluke. When we first computed these
QCD corrections we thought they would dominate because GIM gives an
amplitude with a suppression of $m_t^2/M_W^2$, while our QCD correction
would involve this ratio logarithmically. With $m_t\approx2M_W$ this
argument does not hold water, but remarkably the QCD logarithmic
``corrections'' are still the dominant contribution, because the
function $A$ is small. Remarkably, the prediction we made in 1987 has
changed little by theory refinements and higher order
calculations\cite{Gambino:2001ew}. The reason is that $\alpha_s(m_b)$ is
sufficiently small that perturbation theory works rather well.

Radiative $B$ decays are interesting because they proceed only at one
loop in the standard model. Therefore, new physics at short distances
may in principle compete favorably with the standard model
contribution to the process. To get an idea of how sensitive to new
physics the process is we\cite{Grinstein:1987pu} chose to explore one
of the simplest extensions of the standard model, namely, a
multi-Higgs model. Consider an extension of the standard model with
two Higgs doublets. In order to naturally suppress flavor changing
neutral currents\cite{Glashow:1976nt} it is customary to impose a
symmetry so that all quarks receive their mass from Yukawa couplings
to only one of the two Higgs doublets (``model I''), or so that all
charge~$+2/3$ quarks get their masses from one Higgs doublet and the
charge~$-1/3$ quarks get their masses from the other (``model
II''). Model II is similar to the minimal supersymmetric extension of
the standard model, so it has received more attention. It is easy to
see that the rate $\Gamma(B\to X_s\gamma )$ is strictly larger in model~II than
in the standard model. Figure~8 show the rate in model~II as a
function of $m_t$ for a charged Higgs mass of 100~GeV and with equal
expectation values of the two Higgs doubles ($\tan\beta=1$). Comparing
with Fig.~7 we see that the standard model rate is several times
smaller. Since experiment is consistent with the standard model to
within 20\%, the mass of the charged Higgs in model~II has to be much
larger than 100~GeV. The same is true in minimal supersymmetry, with
the caveat that a light Higgs is allowed if one fine tunes additional
contributions (like higgsinos running in the loop) to cancel the
charged Higgs graph.

\begin{center}
\includegraphics[height=1.9in]{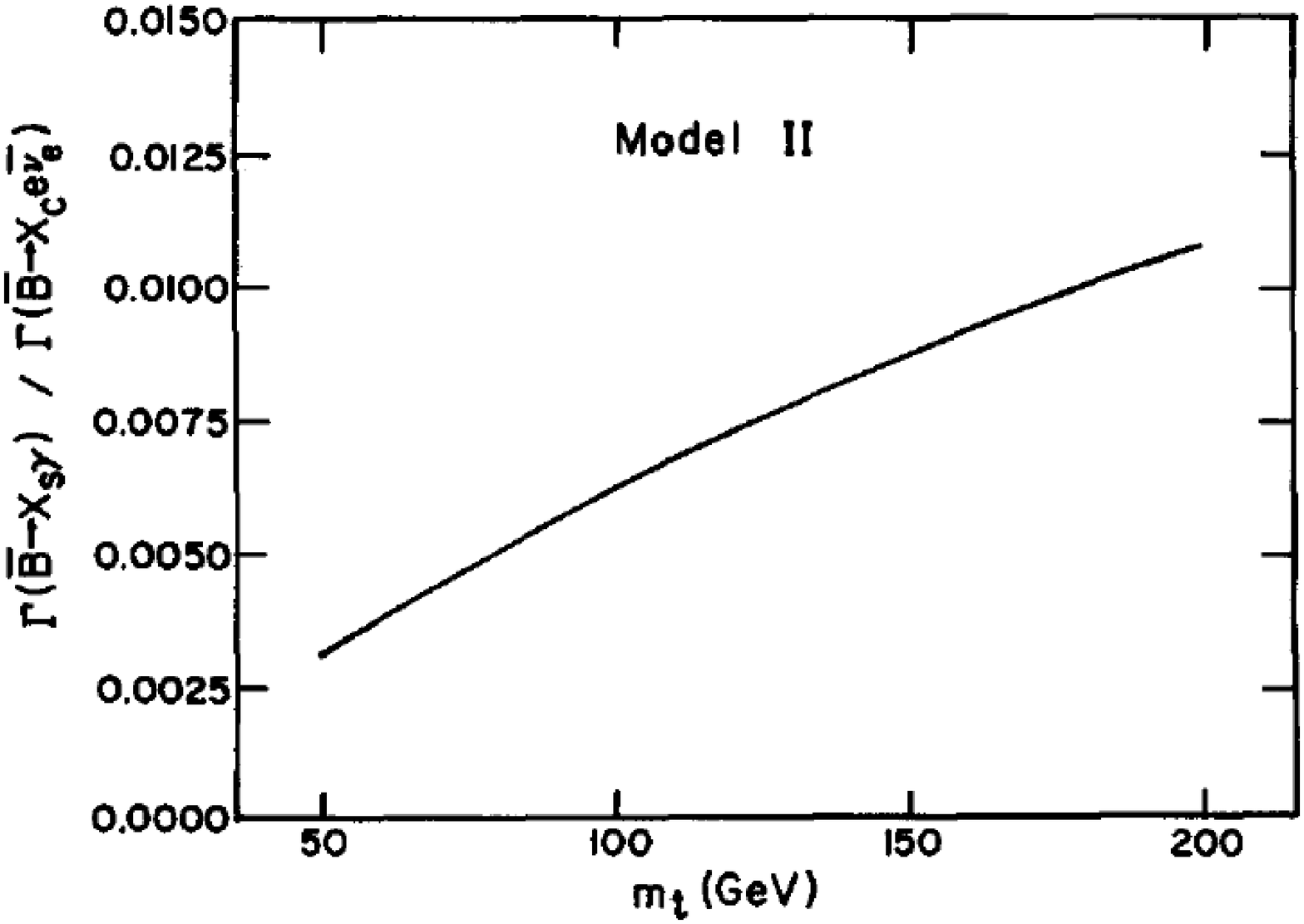}\\
Fig.~8
\end{center}

Many other processes are enhanced by a large top quark mass. This is
demonstrated in Fig.~9\cite{Grinstein:1988me}, which compares the
predicted lepton mass spectrum in the decay $B\to X_se^+e^-$, in units
of the semileptonic decay rate, for $m_t=125$~GeV (lower curve) and
150~GeV (higher curve).

\begin{center}
\includegraphics[height=1.9in]{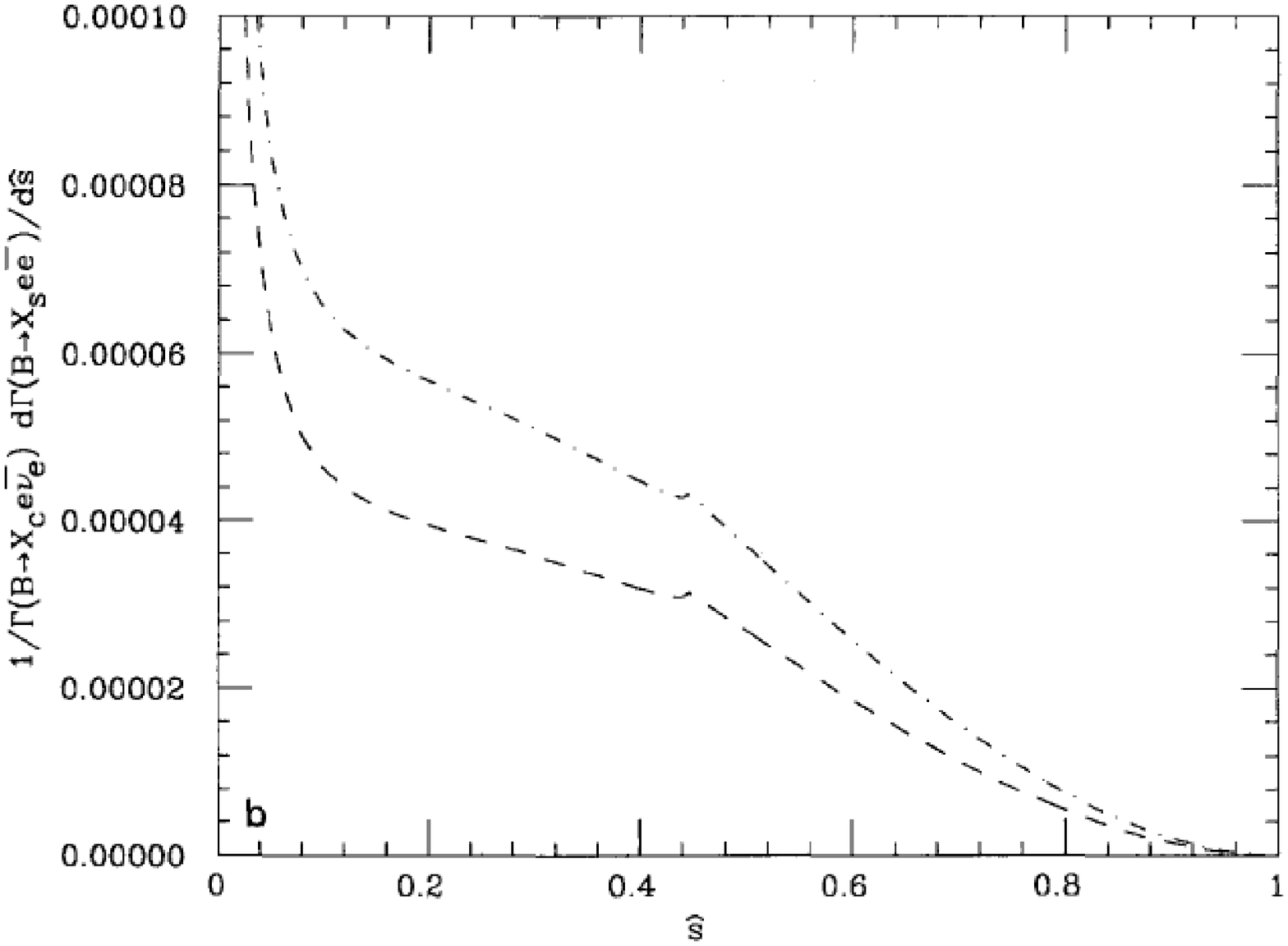}\\
Fig.~9
\end{center}

\section{Precision CKM, HQET}
\subsection{Introduction}
It did not escape the attention of many that the quark model
prediction of the inclusive decay rate, shown in Fig.~2, is dominated
by the lowest two charmed states: the $D$ and $D^*$ mesons. Nussinov
and Wetzel\cite{Nussinov:1986hw} emphasized that this calculation has
nothing to do with the details of the quark model used. They argued
that the $D$, $D^*$ and $B$ wavefunctions are the same, so that when
the $D$ or $D^*$ mesons are not recoiling in the decay the amplitude
is given by the overlap of identical wave-functions which is fixed to
unity for normalized wave-functions. Moreover, the higher resonances
have wavefunctions that are orthogonal to these, so clearly their
contributions are small. The Nussinov-Wetzel argument relied on the
quark-model picture that the meson is a two body system with a
potential binding. Since the $b$ and $c$ are heavy, the reduced mass
(which governs all the dynamics) is the same for both $D$ and $B$
mesons. Moreover, the heavy spin decouples, since the coupling arises
from the magnetic moment which scales inversely with the large
mass. So the wavefunction for the $D^*$ agrees with that of the $D$.

If this were true in QCD then one could begin a program of precision CKM
determination. Enter HQET. Heavy Quark Effective Theory (HQET) is a
calculational method that exploits enhanced symmetries of QCD that
appear when one restricts attention to a very specific sector of the
theory. I stress that HQET {\it is} QCD. This is in contrast with
hadronic models, like the quark potential model, which had been used
to calculate form factors for heavy meson decays.  

The successes of the constituent quark model is indicative of the fact
that, inside hadrons, strongly bound quarks exchange momentum of
magnitude a few hundred MeV. We can think of the typical amount
$\Lambda$ by which the quarks are off--shell in the nucleon as
$\Lambda \approx m_p/3 \approx 330$~MeV. In a heavy hadron the same
intuition can be imported, and again the light quark(s) is(are) very
far off--shell, by an amount of order $\Lambda$.  But, if the mass
$M_Q$ of the heavy quark $Q$ is large, $M_Q \gg \Lambda$, then, in
fact, this quark is almost on--shell. Moreover, interactions with the
light quark(s) typically change the momentum of $Q$ by $\Lambda$, but
change the {\it velocity} of $Q$ by a negligible amount, of the order
of $\Lambda/M_Q \ll 1$. It therefore makes sense to think of $Q$ as
moving with constant velocity, and this velocity is, of course, the
velocity of the heavy hadron.

In the rest frame of the heavy hadron, the heavy quark is practically
at rest.  The heavy quark effectively acts as a static source of
gluons. It is characterized by its flavor and color--$SU(3)$ quantum
numbers, but not by its mass. In fact, since spin--flip interactions
with $Q$ are of the type of magnetic moment transitions, and these
involve an explicit factor of $g_s/M_Q$, where $g_s$ is the strong
interactions coupling constant, the spin quantum number itself
decouples in the large $M_Q$ case.  Therefore, {\it the properties of
heavy hadrons are independent of the spin and mass of the heavy source
of color.}

The HQET is nothing more than a method for giving these observations a
formal basis. It is useful because it gives a procedure for making
explicit calculations. But more importantly, it turns the statement
`$M_Q$ is large' into a {\it systematic} perturbative expansion in
powers of $\Lambda/M_Q$. Each order in this expansion involves QCD to
all orders in the strong coupling, $g_s$. Also, the statement of mass
and spin independence of properties of heavy hadrons appears in the
HQET as approximate internal symmetries of the Lagrangian.

\subsection{Effective Lagrangian and New symmetries}
We shall focus our attention on the calculation of Green functions in
QCD, with a heavy quark line, its external momentum almost on--shell.
The external momentum of gluons or light quarks can be far off--shell,
but not much larger than the hadronic scale $\Lambda$. This region of
momentum space is interesting because physical quantities
---$S$--matrix elements--- live there. And, as stated in the
introduction, we expect to see approximate symmetries of Green
functions in that region which are not symmetries away from it. That
is, these are approximate symmetries of a sector of the $S$--matrix,
but not of the full QCD Lagrangian.

The effective Lagrangian ${\cal L}_{\rm eff}$ is constructed so that it will
reproduce these Green functions, to leading order in $\Lambda/M_Q$. It is
given, for a heavy quark of velocity $v_{\mu}$ ($v^2=1$), 
by\cite{eichtenhill}, 
\begin{equation}\label{eIIi}
 {{\cal L}_{\rm eff}^{(v)} = \bar Q_{v} i v\cdot D  Q_v~,}
\end{equation}
where the covariant derivative is
\begin{equation}\label{eIIii}
 {D_{\mu}= \partial_{\mu} + i g_s A^a_{\mu} T^a~,}
\end{equation}
and the heavy quark field $Q_{v}$ is a Dirac spinor that satisfies the
constraint
\begin{equation}\label{eIIiii}
{\left(\frac{1 + {\not v}}{2}\right) Q_v = Q_v~.}
\end{equation}
In addition, it is understood that the usual Lagrangian ${\cal L}_{\rm light}$
for gluons and light quarks is added to ${\cal L}_{\rm eff}^{(v)}$.

We have introduced an effective Lagrangian ${\cal L}_{\rm eff}^{(v)}$ such
that Green functions $\widetilde G_v(k;q)$ calculated from it agree,
at tree level, with corresponding Green functions $G(p;q)$ in  QCD
to leading order in the large mass
\begin{equation}
\label{eIIv}
{
G(p;q) = \widetilde G_v (k;q) + {\cal O} \left({\Lambda / M_Q}\right)
\qquad\qquad\hbox{(tree level)}~.
}
\end{equation}
Here, $\Lambda$ stands for any component of $k_{\mu}$ or of the $q$'s,
or for a light quark mass, and $p=M_Q v+k$. It is straightforward to
verify Eq.~(\ref{eIIv}). 

Beyond tree level the corrected version is still close in form to
this\cite{Grinstein:1990mj},
\begin{equation}\label{eIIvi}
G(p;q;\mu) = C({M_Q / \mu}, g_s) \widetilde G_{v}(k;q;\mu) 
+ {\cal O}\left({\Lambda /  M_Q}\right)~. 
\end{equation}
The Green functions $G$ and $\widetilde G_{v}$ are renormalized, so
they depend on a renormalization point $\mu$. The function 
$C$ is independent of momenta or light quark masses: it is independent
of the dynamics of the light degrees of freedom. It is there because
the left hand side has some terms which grow logarithmically with the
heavy mass, $\ln( M_Q/\mu)$. The beauty of Eq.~(\ref{eIIvi}) is that {\it
all of the logarithmic dependence on the heavy mass factors out.}
Better yet, since $C$ is dimensionless, it is a function of
the ratio $M_Q/\mu$ only, and not of $M_Q$ and $\mu$
separately. (Actually, additional $\mu$ dependence is implicit in
the definition of the renormalized coupling constant $g_s$.  This
reflects itself in the explicit form of $C$). To find the
dependence on $M_Q$ it suffices to find the dependence on $\mu$.  This
in turn is dictated by the renormalization group equation.

It is appropriate to think of the HQET as a factorization theorem,
stating that, in the large $M_Q$ limit, the QCD Green functions
factorize into a universal function of $M_Q$, 
$C({M_Q / \mu}, g_s)$, which depends on the short distance
physics only, times a function that contains all of the information
about long distance physics and is independent of $M_Q$, and can be
computed as a Green function of the HQET lagrangian.

The Lagrangian for $N$ species of heavy quarks, all with velocity $v$, is
\begin{equation}\label{hqeteffl}
{\cal L}_{\rm eff}^{(v)} = \sum^N_{j=1} \bar Q_v^{(j)} \; i v \cdot D\; Q^{(j)}_v~.
\end{equation}
This Lagrangian has a $U(N)$ symmetry{\cite{isgurwisea,volosh}}. The
subgroup $U(1)^N$ corresponds to flavor conservation of the strong
interactions, and was a good symmetry in the original theory. The
novelty in the HQET is then the non-Abelian nature of the symmetry
group. This leads to relations between properties of heavy hadrons
with different quantum numbers. Please note that these will be
relations between hadrons of a given velocity, even if of different
momentum (since typically $M_{Q_i} \neq M_{Q_j}$ for $i \neq j$).  Including
the $b$ and $c$ quarks in the HQET, so that $N=2$, we see that the $B$
and $D$ mesons form a doublet under flavor--$SU(2)$.

This {\sl flavor--$SU(2)$} is an approximate symmetry of QCD. It is a good
symmetry to the extent that 
\begin{equation}\label{largemass}
{
m_c \gg \Lambda\qquad\hbox{and}\qquad  m_b \gg \Lambda ~.}
\end{equation}
These conditions can be met even if $m_b - m_c \gg \Lambda$.  This is in
contrast to isospin symmetry, which holds because $m_d-m_u \ll \Lambda$.

In atomic physics  this symmetry implies
the equality of chemical properties of different isotopes of an element.

{\sl  Spin -- $SU(2)$:}
The HQET Lagrangian involves only two components of the spinor $Q_v$.
Recall that 
\begin{equation}
\label{projonQ}
\left(\frac{1-{\not v}}{2}\right) Q_v = 0.
\end{equation}
The two surviving components enter the Lagrangian diagonally, {\it
i.e.}, there are no Dirac matrices in $\mathcal{L}_{\text{eff}}$ in
Eq.~(\ref{eIIi}).  Therefore, there is an $SU(2)$ symmetry of this
Lagrangian which rotates the two components of $Q_v$ among
themselves\cite{isgurwisea,eichtenfeinberglepagethacker}.

Please note that this ``spin''--symmetry is actually an {\sl internal}
symmetry. That is, for the symmetry to hold no transformation on the
coordinates is needed, when a rotation among components of $Q_v$ is
made. On the other hand, to recover Lorentz covariance, one does the
usual transformation on the light--sector, including a Lorentz
transformation of coordinates and in addition a Lorentz transformation
on the velocity $v_\mu$. A spin--$SU(2)$ transformation can be added to
this procedure, to mimic the original action of Lorentz
transformations.

\subsection{Exclusive Semileptonic Decays}
The symmetries in HQET are sufficient to give us the matrix elements
for semileptonic $B$ decay in terms of one undetermined ``Isgur-Wise''
function $\xi(v\cdot v')$:
\begin{multline}
\label{eq-IW}
\vev{D(v')\vert V_\mu\vert \bar B(v)}=
\xi(v\cdot v')(v_\mu+v'_\mu), \\
\vev{D^\ast(v')\epsilon\vert V_\mu- A_\mu\vert\ \bar B(v)}=
\xi(v\cdot v')[i\epsilon_{\mu\nu\lambda\sigma}\epsilon^{\ast\nu}v^\lambda v^\sigma\\
+\epsilon^\ast_\mu(1+v\cdot v')-v'_\mu\epsilon^\ast\cdot v].
\end{multline}
Moreover, the Isgur-Wise function satisfies a normalization condition,
$\xi(1)=1$. It is easy to see where this comes from. The forward
scattering amplitude of the $B$ meson by the $b$-current, $\bar b \gamma^\mu
b$, is normalized by charge conservation (just like electromagnetic
charge form factors). An HQ-flavor $SU(2)$ transformation relates this
to the first matrix element in (\ref{eq-IW}) and a further HQ-spin
$SU(2)$ transformation relates it to the second line in
(\ref{eq-IW}). This is remarkable: six unknown form factors are given
in terms of one, which, in addition, is known at one kinematic point!

A remarkable theorem by Luke\cite{Luke:1990eg} states that even after
including corrections of order $1/M_Q$ some form factors are still
normalized. The remaining irreducible uncertainty in the determination
of $|V_{cb}|$ is of order $(\Lambda/2m_c)^2\sim0.01$, so this should give a
determination with precision of a few per-cent. The measurement is
complicated by the fact that the decay rate vanishes at $ v\cdot v'=1$,
so the measurement needs to be extrapolated. Fortunately, QCD
restricts significantly the extrapolation\cite{Boyd:1997kz} so little
uncertainty is introduced.

\subsection{Inclusive Semileptonic Decays and Duality}
Quark-hadron duality, the imprecise statement that a quantity can be computed
directly at the parton level if it is inclusive enough and the energy
involved is large enough, had long been thought to hold for decay rates of
heavy mesons. However, in fact, the statement
\begin{equation}
\frac{d\Gamma(B\to X_q e \bar\nu)}{dE_e dm^2_{e\bar\nu}} \stackrel{?}{=}
\frac{d\Gamma(b\to q e \bar\nu)}{dE_e dm^2_{e\bar\nu}},
\end{equation}
with $q=u$ or $c$, is generally not correct. But if we smear over
the electron energy things work out\cite{Bjorken:1990hs,Chay:1990da}:
\begin{equation}
\label{eq:duality}
\left\langle \frac{d\Gamma(B\to X_q e \bar\nu)}{dE_e dm^2_{e\bar\nu}} 
\right\rangle_f
=
\left\langle \frac{d\Gamma(b\to q e \bar\nu)}{dE_e dm^2_{e\bar\nu}}\right\rangle_f\left(1 +
\mathcal{O}\left(\frac{1}{m_b^2}\right)\right)~,
\end{equation}
where the smearing is defined by
\begin{equation}
\vev{g}_f \equiv\int dE f(E)g(E)
\end{equation}
with $f$ a smooth function.

The result is obtained by simultaneous short distance (OPE) and heavy
quark mass expansions of a Green's function, the physical rate given
by its imaginary part (discontinuity across the cut). One therefore
gets, in addition, a means for systematically improving the expansion,
order by order in $1/m_b$. We have indicated this in
Eq.~(\ref{eq:duality}), which shows the remarkable
result\cite{Chay:1990da} that there are no corrections at first order
in $1/m_b$. Moreover, by choosing the function $f$ appropriately one
can find new sum rules. These allow, for example, experimental
determination of the unknown parameters that appear at order
$1/m_b^2$\cite{Manohar:1993qn} by computing moments of the
spectrum\cite{Falk:1995me}.

\subsection{Nailing down $|V_{cb}|$ and $|V_{ub}|$}
For semileptonic decays to charm we have seen that our theoretical
understanding of both exclusive and inclusive widths is solid. At the
time of the Colima conference HFAG was quoting remarkable agreement in
the determination of $V_{cb}$ by both means:
\begin{equation}
|V_{cb}| =\begin{cases}
(41.9\pm1.1_{\rm exp}\pm1.8_{\xi(1)})\times10^{-3}
 &\text{exclusive,}\\
(41.2\pm0.7_{\rm exp}\pm0.6_{\rm th})\times10^{-3}
 &\text{inclusive.}
\end{cases}
\end{equation}
The updated results can be found in HFAG's website\cite{HFAG}.

The prospect of making a precise determination of $|V_{ub}|$ is not as
bright. There is no simple way of determining the form factors. One
can either rely on the prospect of future precise lattice
computations, or on using symmetry to fix the form factors
indirectly. It may be possible to determine $|V_{ub}|$ to few per-cent
accuracy by comparing the rates for $B\to \rho\ell\nu$, $B\to K^*\ell\bar\ell $,
$D\to\rho\ell\nu $ and $D\to K^*\ell\nu $\cite{Grinstein:2004vb}. 
The endpoint of the electron/muon energy spectrum in
$b\to u\ell\nu $ cannot be reliably described theoretically: the OPE/HQET
expansion breaks down (not enough smearing in
Eq.~(\ref{eq:duality}). One may limit the charm
contamination by restricting other kinematic variables. Decay to charm
is not allowed for final state hadronic invariant mass $m_X<m_D$, or
for final state lepton pair invariant mass square
$q^2>(m_B-m_D)^2$. The best method will most likely involve carving
out a region of $m_X$ {\it vs} $q^2$ space that minimizes theoretical
errors while keeping the charm background under control\cite{Bauer:2000xf}.

\section{Factorization, LEET and SCET}
Roughly speaking, factorization in $B\to D\pi$ means that the following
holds,
\begin{equation}
\label{eq:factorization}
\vev{D\pi|(\bar ud)_{V-A}(\bar bc)_{V-A}|B}\approx 
\vev{D|(\bar bc)_{V-A}|B}
\vev{\pi|(\bar ud)_{V-A}|0},
\end{equation}
and can be tested experimentally by comparing
$\Gamma(B\to D\pi)$ and $f_\pi^2d\Gamma(B\to D\ell\nu)/dM^2_{\ell\nu}$ at
$M^2_{\ell\nu}=m_\pi^2$. Factorization holds in the large $N_c$ expansion of
QCD to leading order in $1/N_c$, but this explanation of factorization
is a bit too democratic: it gives that factorization holds with the same
accuracy for light meson decays, where it is known to fail
badly. Bjorken suggested\cite{Bjorken:1988kk} that factorization may
hold by  color transparency. One can quantify and systematize
Bjorken's ideas  by showing that factorization holds to all orders in
QCD perturbation theory. One needs to find a proper expansion
parameter such that in leading order gluons originating from the $B/D$
system do not couple to the $\pi$ system. 

One can easily accomplish this\cite{Dugan:1990de}. The idea is to
expand in the large energy, $E=(m_B^2-m_D^2)/2m_B$, released to the
light meson. We treat $b$ and $c$ as heavy quarks, expanding in
$1/m_b$ while keeping $m_c/m_b$ fixed. Then the energy $E$ scales with
$m_b$ while the relativistic $\gamma$-factor of the recoiling $D$
meson is fixed and small, $\gamma=v\cdot v'=1/2[(m_D/m_B)+(m_B/m_D)]\approx1.6
$. Since the velocity of the recoiling $D$ is small, the spectator
quark (the light quark bound in the $B$ meson) only requires a soft
kick to be incorporated into the final $D$ meson. The two light quarks
from the $W$-vector-boson of the weak interaction must produce a
single light meson. So the $\bar ud$ quark pair is produced with
large energy and low invariant mass. This means they travel in the same
light-like direction, and therefore their color charge cancels. This
can be made quantitatively precise by introducing an effective theory
that systematizes an expansion in $1/E$. The LEET\cite{Dugan:1990de}
is an effective theory that includes the interactions of soft gluons
with the relativistic low mass quark-antiquark pair. Decoupling of the
pair from soft gluons is trivial to show in the LEET. Hard gluons do
not decouple, but their effects are suppressed by $\alpha_s(m_b)$, so
corrections to factorization come in both at order $1/E$ and
$\alpha_s(m_b)$. This argument misses the possible effects of collinear
gluons, but these are easily incorporated in SCET, an effective theory
with both soft and collinear gluons\cite{Bauer:2000ew}.

The LEET/SCET justification for factorization in $B\to D\pi$ is
interesting in several respects. First, it is patently different from
large $N_c$ arguments. It applies only to decays of a heavy meson to a
heavy meson plus a light one. This is good. Empirically, factorization
does not hold in $D$ decays. It does not hold in decays of $B$ to two
light mesons, but this case is complicated by the presence of
penguins. In  LEET/SCET factorization the spectator quark needs only a
soft kick to join the final state. When the final state has two light
mesons the spectator quark needs a hard kick, but hard kicks break
factorization (in the sense of Eq.~(\ref{eq:factorization})
above). To be sure, there is a factorizable contribution to the
amplitude, but it is by no means dominant. Second, although the
argument was inspired by Bjorken's color transparency ideas, it is by
no means equivalent. Not only does the effective theory provide a
systematic approach to the study of factorization (or rather, to
violations to factorization) but also it does not justify factorization in
some processes for which color transparency applies, like $B\to J/\psi
K$. And third, it is predictive and therefore testable. For example,
it predicts that the whole amplitude for $B^0\to D^0\pi^0$ is suppressed by
the amount by which factorization is violated,
\begin{equation}
\frac{\mathcal{M}(B^0\to D^0\pi^0)}{ \mathcal{M}(B^0\to
  D^-\pi^+)}\sim\frac{\text{1 GeV}}{m_B}.
\end{equation}
Experimentally, $\text{Br}(B^0\to D^0\pi^0)/ \text{Br}(B^0\to D^-\pi^+)
=(1.0\pm0.3)\times10^{-1}$ in agreement with expectations. 

LEET/SCET factorization in $B\to DX$ holds provided the hadronic system
$X$ has small invariant mass. This is verified experimentally in the
case when $X$ is a single resonance, $X=\pi,\rho, a_1$. As the invariant
mass of $X$ increases and eventually scales with $m_b$ the LEET/SCET
argument fails. This could be tested with $X$ a multi-pion final
state, by measuring the rate as a function of its invariant mass,
$m_X$\cite{Dugan:1990de}. There is evidence that factorization holds
even as a function of $m_X$\cite{Ligeti:2001dk}. There is no
convincing explanation for this.

\section{Duality, CP violation, Conclusions }
Time and space limitations preclude me from discussing other subjects
which are very interesting. I would like to at least mention two of
them:

{\it Quark-Hadron Duality:} I explained above how quark-hadron duality
works in semileptonic decays. There is no analogous treatment for
purely hadronic decays and hence for lifetimes. The problem is that
there is no external kinematic variable that allows us to consider a
Green function, rather than a decay rate. The Green function can be
studied at non-physical momenta and is amenable to an OPE. In the
absence of an OPE we are left with educated guesses. We can study the
question in soluble models. In $1+1$~QCD in the large $N_c$ limit one
can compute exactly the total width of a heavy meson. It is found that
generally the width (and hence the lifetime) differs from the local
quark-hadron duality prediction at order
$1/m_b$\cite{Grinstein:1997xk}, to be contrasted with the $1/m_b^2$
corrections in the semileptonic case. Moreover and also in contrast
with the semileptonic case, there is no systematic theory of
these corrections. 

{\it CP violation:} The effort to precisely determine the elements of
the CKM matrix complements the measurement of the angles of the
unitarity triangle through CP violating asymmetries at
$B$-factories. The idea is to over-constrain the triangle in an effort
to ferret out any hint of new physics hiding under the surface. Bigi
and Sanda observed that CP asymmetries through the interference of
mixing and decay of neutral $B$ mesons to states that are CP
eigenstates determine unitarity angles cleanly, without contamination
from unknown hadronic matrix elements\cite{Bigi:1981qs}. This works
extremely well for determinations of $\sin(2\beta)$ through, for example,
$B\to J/ \psi K_S$. One of the challenges of modern and future $B$
factories is to determine $\alpha$ and $\gamma$ as well. A very unwelcome
surprise came several year after Bigi and Sanda's ground-breaking
paper when it was first realized that the determination of $\sin(2\alpha)$
in, say, $B\to\pi\pi$, could be compromised by unknown hadronic matrix
elements because the effects of penguin diagrams could be
significant\cite{Grinstein:1989df}.

{\it Conclusions.} I learned about the discovery of charm, the third
generation of leptons and then of $b$ quarks from  Scientific American
articles when I was in High School and then in College. It has been a
fabulous opportunity to contribute to the understanding of these
beasts throughout the time over which this field has evolved. The
story is similar to that of electroweak-physics, and one could say
that the $B$ meson is to CKM as the $Z$ vector-boson is to the
SM. With the big caveat that $B$ physics is still furiously evolving
and may still hold some surprises.
\bigskip 

{\it Acknowledgments.} I am very grateful to the Divisi\'on de
Part\'\i culas y Campos of the Sociedad Mexicana de F\'\i sica for conferring
upon me the Medal 2003, and for the opportunity to give this talk at
the {\it IX Mexican Workshop on Particles and Fields}. I would like to
thank Prof.\ Myriam Mondrag\'on for the nomination to the Medalla~2003, and
my former mentors at CINVESTAV for their support, Miguel
Angel P\'erez, Augusto Garc\'\i a and Arnulfo Zepeda in particular. Thanks
also to my colleagues at the Facultad de Ciencias of the Universidad
the Colima for their hospitality. This work is supported in part by a
grant from the Department of Energy under Grant DE-FG03-97ER40546.

\end{multicols}
\medline
\begin{multicols}{2}
\newcommand{\prl}[3]{Phys. Rev. Lett. {\bf #1}, #3 (#2).}
\newcommand{\pr}[3]{Phys. Rev. {\bf #1}, #3 (#2).}
\newcommand{\pl}[3]{Phys. Lett. B {\bf #1}, #3 (#2).}
\newcommand{\np}[3]{Nucl. Phys. B. {\bf #1}, #3 (#2).}
\newcommand{\sjnp}[3]{Sov. J. Nucl. Phys. {\bf #1}, #3 (#2).}

\end{multicols}
\end{document}